\documentclass[aip,jcp,amsmath,amssymb,graphicx,twocolumn,superscriptaddress, reprint]{revtex4-1} 
\usepackage{svg}
\usepackage{graphicx}
\usepackage{dcolumn}
\usepackage{bm}
\usepackage{comment}
\usepackage{multirow}
\usepackage{array}
\usepackage{tabularx}
\usepackage{amsmath}
\usepackage{xcolor}
\usepackage[normalem]{ulem}
\usepackage{hyperref}
\usepackage{soul}
\setulcolor{red}
\usepackage{booktabs}
\usepackage{siunitx}

\draft 

\AtBeginDocument{%
  \heavyrulewidth=.08em
  \lightrulewidth=.05em
  \cmidrulewidth=.03em
  \belowrulesep=.65ex
  \belowbottomsep=0pt
  \aboverulesep=.4ex
  \abovetopsep=0pt
  \cmidrulesep=\doublerulesep
  \cmidrulekern=.5em
  \defaultaddspace=.5em
}

\begin{document}

\makeatletter
\def\@email#1#2{%
 \endgroup
 \patchcmd{\titleblock@produce}
  {\frontmatter@RRAPformat}
  {\frontmatter@RRAPformat{\produce@RRAP{*#1\href{mailto:#2}{#2}}}\frontmatter@RRAPformat}
  {}{}
}%
\makeatother
\title{Electronic structure of zaykovite Rh$_3$Se$_4$, prediction and analysis of physical properties of related materials: Pd$_3$Se$_4$, Ir$_3$Se$_4$, and Pt$_3$Se$_4$} 

\author{Leonid S. Taran}
\email{leonidtaran97@gmail.com}
\affiliation{M. N. Mikheev Institute of Metal Physics, Ural Branch of Russian Academy of Sciences,\\620137 Yekaterinburg, Russia}

\author{Sergey V. Eremeev}%
\affiliation{Institute of Strength Physics and Materials Science of Siberian Branch of Russian Academy of Sciences,\\634055 Tomsk, Russia}

\author{Sergey V. Streltsov}
\affiliation{M. N. Mikheev Institute of Metal Physics, Ural Branch of Russian Academy of Sciences,\\620137 Yekaterinburg, Russia}

\date{\today}

\begin{abstract}
In this work, we explore the electronic properties and chemical bonding in the recently discovered mineral zaykovite, the first natural rhodium selenide Rh$_3$Se$_4$. We comprehensively studied the bulk electronic structure, hybridization of rhodium and selenium orbitals, and the influence of spin-orbit interaction on the electronic spectrum, as well as inspected its topological properties. Besides, we investigated the surface electronic structure of zaykovite and revealed the anisotropic Rashba-type spin splitting in the surface states.
In addition, using calculations of the phonon spectra and enthalpy of formation we predicted the family of similar selenides based on other $4d$ and $5d$ transition metals such as Ir, Pd, and Pt. The structural and electronic properties of these materials are discussed.

\end{abstract}


\maketitle 

\section{INTRODUCTION}
While hundreds of thousands of crystal structures of inorganic materials are known to the date~\cite{Zagorac2019}, there are only 6062 official minerals (as of July 2024), with $\sim 100$ new minerals being discovered each year~\cite{Olds2024}. Moreover, typically these new minerals turns out to have a rather complex chemical formula.  The mineral zaykovite, recently discovered in the Kazan gold placer~\cite{Belogub2023}, has not only a simple formula Rh$_3$Se$_4$ (of course natural samples include different types of impurities and are slightly off-stoichiometric), but also turns out to be the first known natural rhodium selenide. Moreover, this mineral contains two heavy elements, which are known to have rather strong spin-orbit coupling (SOC), and therefore its electronic structure can potentially exhibit both non-trivial band topology and/or Rashba-type spin splitting when translation symmetry is broken, i.e., on a surface.

Zaykovite was found in a continuous series of solid solutions with structurally similar~\cite{Belogub2023} and well known kingstonite Rh$_3$S$_4$~\cite{Stanley2005,Dieguez2009}. While the crystal structure of Rh$_3$Se$_4$ has been refined, its physical and chemical properties remain poorly studied. Only investigations of synthesized Rh$_3$Se$_4$ nanoparticles and heterostructures for catalysis applications such as oxygen reduction reaction (ORR) \cite{Pan2021_ORR,Golubovic2024} and hydrogen evolution reaction (HER) \cite{Pan2021_HER}  have been earlier reported. This work aims to investigate the electronic structure of zaykovite using the first-principles calculations, which can further help the chemical-physics community in the search for more reliable electrocatalyst and material scientists in a qualitative study of similar minerals. The composition of the natural zaykovite crystals contains platinum, palladium, and iridium impurity atoms, substituting the rhodium atoms in a small ratio~\cite{Belogub2023}. In this regard, it is of interest to consider hypothetical new selenides in which all rhodium atoms are replaced by the impurity atoms. We demonstrate that
$X_3$Se$_4$ family ($X=\mathrm{Pt,Pd,Ir}$) compounds are indeed chemically and structurally stable and study their electronic properties. Natural zaykovite also contains sulfur impurities on the selenium sublattice, but consideration of sulfides is beyond the scope of our work.

\begin{figure}[h!]
\centering
\includegraphics[width=1\columnwidth]{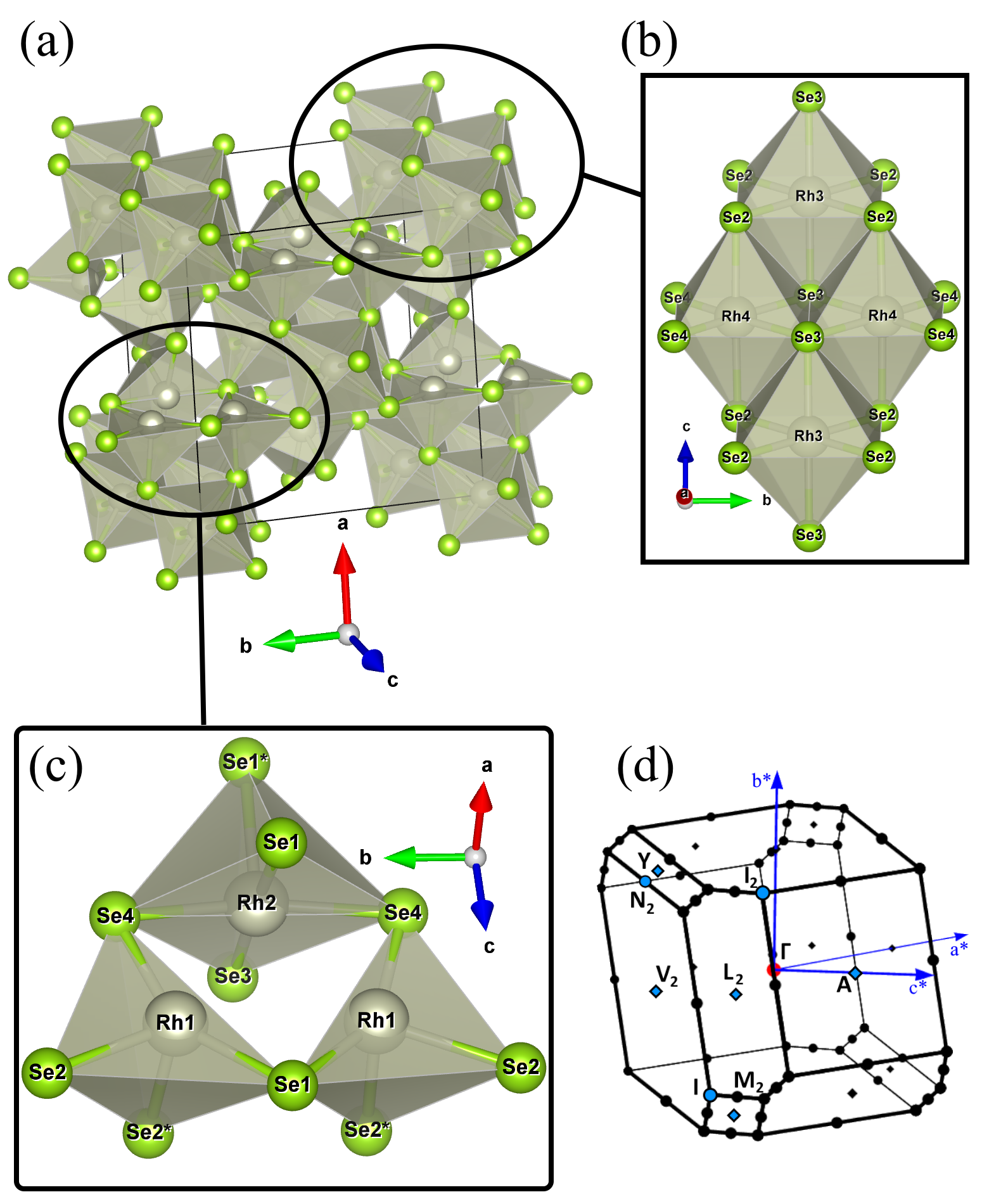}
    \caption{(a) Polyhedral representation of the Rh$_3$Se$_4$ crystal structure. Olive-yellow polyhedra correspond to the (b) octahedra (Rh3 and Rh4) chains, (c) Rh2 pyramids and Rh1 tetrahedrons. Rh1-Se2* and Rh2-Se1* bond distances are different from the corresponding bonds without asterisk, see Table 1; (d) Brillouin zone for the primitive cell with high-symmetry points.}
    \label{fig:Rh3Se4_Structure}
\end{figure}

\section{Methods}
All calculations were performed using the Perdew-Burke-Ernzerhof version of the generalized gradient approximation (GGA) \cite{Perdew1997} employing the \textsc{vasp} code \cite{Kresse1996}. In order to account for dispersion corrections, the DFT-D3 method with Becke-Johnson damping was used \cite{Grimme2010,Grimme2011}. The cutoff energy for the plane-wave basis was set to 280 eV. Stopping criterion for the electronic self-consistency was 10$^{-7}$ eV. The Brillouin zone integration was carried out over $3\times 3 \times 4$ Monkhorst-Pack mesh \cite{Monkhorst1976}. Specified Wigner-Seitz radii for rhodium, iridium, palladium, platinum and selenium are 1.402, 1.423, 1.434, 1.455 and 1.164 \AA ~respectively. A series of calculations including spin-orbit coupling (GGA+SOC) have also been carried out. All considered crystal structures were subjected to a full relaxation procedure (atomic positions, cell shape and volume) by the conjugate gradient algorithm \cite{Press1986}. 
A force tolerance criterion for convergence of atomic positions was set to 10$^{-3}$ eV/\AA, while convergence criterion for the total energy was chosen to be 10$^{-6}$ eV. For the dynamic stability investigation, the first-principles phonon calculations using phonopy were performed\cite{Togo2023,Togo2023-2}. The onsite Coulomb interaction was taken into account via a rotationally invariant DFT+$U$ approach after Dudarev \textit{et al.} \cite{Dudarev1998}. To visualize and analyse chemical bonding, the Crystal Orbital Hamiltonian Populations (COHP) method \cite{Dronskowski1993} in the plane-wave realization (projected COHP, pCOHP)~\cite{Deringer2011} was performed employing the \textsc{lobster} package \cite{Maintz2016,Nelson2020}. The presented atomic structures were visualized with {\sc vesta} \cite{VESTA}. 

\section{$\mathrm{Rh}_3\mathrm{Se}_4$: crystal and electronic structure}

\subsection{Crystal structure}
\begin{table}[t!]
    \centering
    \caption{Unit cell parameters and Rh-Se bond lengths ($d_\mathrm{NN}$) presented in Fig.\ref{fig:Rh3Se4_Structure} for the experimental zaykovite and relaxed cells using GGA and GGA-D3 (which takes into account van-der-Waals-dispersion energy correction) methods. Bonds with asterisks differ in length from their counterparts without an asterisk.}
    \begin{ruledtabular}
    \begin{tabular}{cccc}
        Structure & Experimental\cite{Belogub2023} & GGA & GGA-D3 \\ 
        \midrule 
        $a$ (\AA) & 10.877 & 10.990  & 10.886  \\ 
        $b$ (\AA) & 11.192 & 11.525  & 11.395  \\ 
        $c$ (\AA) & 6.480 & 6.602 & 6.527  \\ 
        $\alpha$ (deg) & 90 & 90  & 90  \\ 
        $\beta$ (deg) & 108.887 & 107.823  & 107.920  \\ 
        $\gamma$ (deg) & 90 & 90  & 90  \\ 
        $V$ (\AA$^3$) & 746.331 & 796.124 & 770.345 \\ 
        \midrule
        \multicolumn{4}{c}{$d_\mathrm{NN}$ (\AA), tetrahedron environment} \\
        \midrule 
        Rh1-Se1 & 2.359 & 2.417 & 2.391 \\ 
        \: Rh1-Se2* & 2.518 & 2.525 & 2.493 \\ 
        Rh1-Se2 & 2.480 & 2.479 & 2.459  \\ 
        Rh1-Se4 & 2.349 & 2.375 & 2.359  \\
        \midrule 
        \multicolumn{4}{c}{$d_\mathrm{NN}$ (\AA), pyramidal environment} \\
        \midrule 
        \: Rh2-Se1* & 2.440 &  2.499 &  2.455 \\ 
        Rh2-Se1 & 2.414 & 2.488 & 2.475 \\ 
        Rh2-Se3 & 2.431 & 2.484 & 2.448 \\ 
       \qquad  Rh2-Se4\,($\times 2$) & 2.350 & 2.405 & 2.387 \\
        \midrule 
        \multicolumn{4}{c}{$d_\mathrm{NN}$ (\AA), octahedral environment} \\
        \midrule 
       \qquad  Rh3-Se2\,($\times 4$) & 2.493 & 2.548 & 2.517 \\ 
       \qquad  Rh3-Se3\,($\times 2$) & 2.452 & 2.504 & 2.478 \\
        \midrule 
        \multicolumn{4}{c}{$d_\mathrm{NN}$ (\AA), octahedral environment} \\
        \midrule 
       \qquad  Rh4-Se2\,($\times 2$) & 2.473 & 2.520 & 2.498 \\ 
       \qquad  Rh4-Se3\,($\times 2$) & 2.434 & 2.487 & 2.463 \\
       \qquad  Rh4-Se4\,($\times 2$) & 2.388 & 2.434 & 2.406 \\ 
    \end{tabular}
    \end{ruledtabular}
    \label{Tab:struct}
\end{table}

\begin{table}[t!]
    \centering
    \caption{Atomic coordinates for Rh$_3$Se$_4$ obtained by the relaxation of the crystal structure in GGA-D3.}
    \begin{ruledtabular}
    \begin{tabular}{lccc}
        Site & $x$ & $y$ & $z$ \\
        \cmidrule(lr){1-4}
        \,Rh1 (8\textit{j}) & 0.36669 & 0.14558 & 0.95466  \\ 
        \,Rh2 (4\textit{i}) & 0.35215 & 0 & 0.56214 \\ 
        \,Rh3 (2\textit{a}) & 0 & 0 & 0 \\ 
        \,Rh4 (4\textit{h}) & 0 & 0.16053 & 0.5 \\ 
       \,Se1 (4\textit{i}) & 0.41518 & 0 & 0.23914 \\ 
       \,Se2 (8\textit{j}) & 0.12905 & 0.15742 & 0.88954 \\ 
       \,Se3 (4\textit{i}) & 0.11803 & 0 & 0.39084 \\ 
       \,Se4 (8\textit{j}) & 0.35729 & 0.20809 & 0.60564 \\
    \end{tabular}
    \end{ruledtabular}
    \label{table:Rh3Se4AtomCoords}
\end{table}
The initial crystal structure was taken from Ref.~ [\onlinecite{Belogub2023}], where the lattice parameters were obtained using powder X-ray diffraction. This structure belongs to the monoclinic crystal system with the \textit{C2/m} space group, and its lattice parameters and interatomic distances are presented in the column labeled ``experimental'' in the Table~\ref{Tab:struct}.  In ideal case [only rhodium and selenium atoms, see Fig.~\ref{fig:Rh3Se4_Structure}(a)], there are six formula units of Rh$_3$Se$_4$ per unit cell, where each formula unit contains four crystallographycally inequivalent Rh and Se atoms (Rh1 -- Rh4 and Se1 -- Se4). Ribbons of edge-sharing Rh3 and Rh4 octahedra [RhSe$_6$] are parallel to the \textit{c}-axis [Fig.~\ref{fig:Rh3Se4_Structure}(b)], and between them are quadrangular pyramids of Rh2 [RhSe$_5$] and Rh1 tetrahedra [RhSe$_4$] [Fig.~\ref{fig:Rh3Se4_Structure}(c)].

Since natural minerals include a significant amount of impurities and inclusions of other atoms (up to 20\% of the platinum and 30\% of sulfur instead of rhodium and selenium), we performed the full relaxation of the crystal structure. The lattice constants and interatomic distances after optimizations obtained with GGA and GGA-D3 (GGA+DFT-D3) are presented in Table~\ref{Tab:struct}. As can be seen, the equilibrium volume determined in both cases overestimates the experiment by about 6.7~\% and 3.2~\%, respectively. This could well be explained by the divergence of compositional variation in the natural crystal, as sulfur made up the majority of the impurities and was replaced by selenium. On the other hand, account of the London-dispersion correction by the GGA-D3 method improves the situation reducing the cell volume considerably. Therefore, all further calculations are carried out with the GGA-D3 relaxed crystal structure, the equilibrium atomic coordinates of which are given in the Table~\ref{table:Rh3Se4AtomCoords}.

\subsection{Bulk electronic structure}
\begin{figure}[t!]
\centering
\includegraphics[width=1\columnwidth]{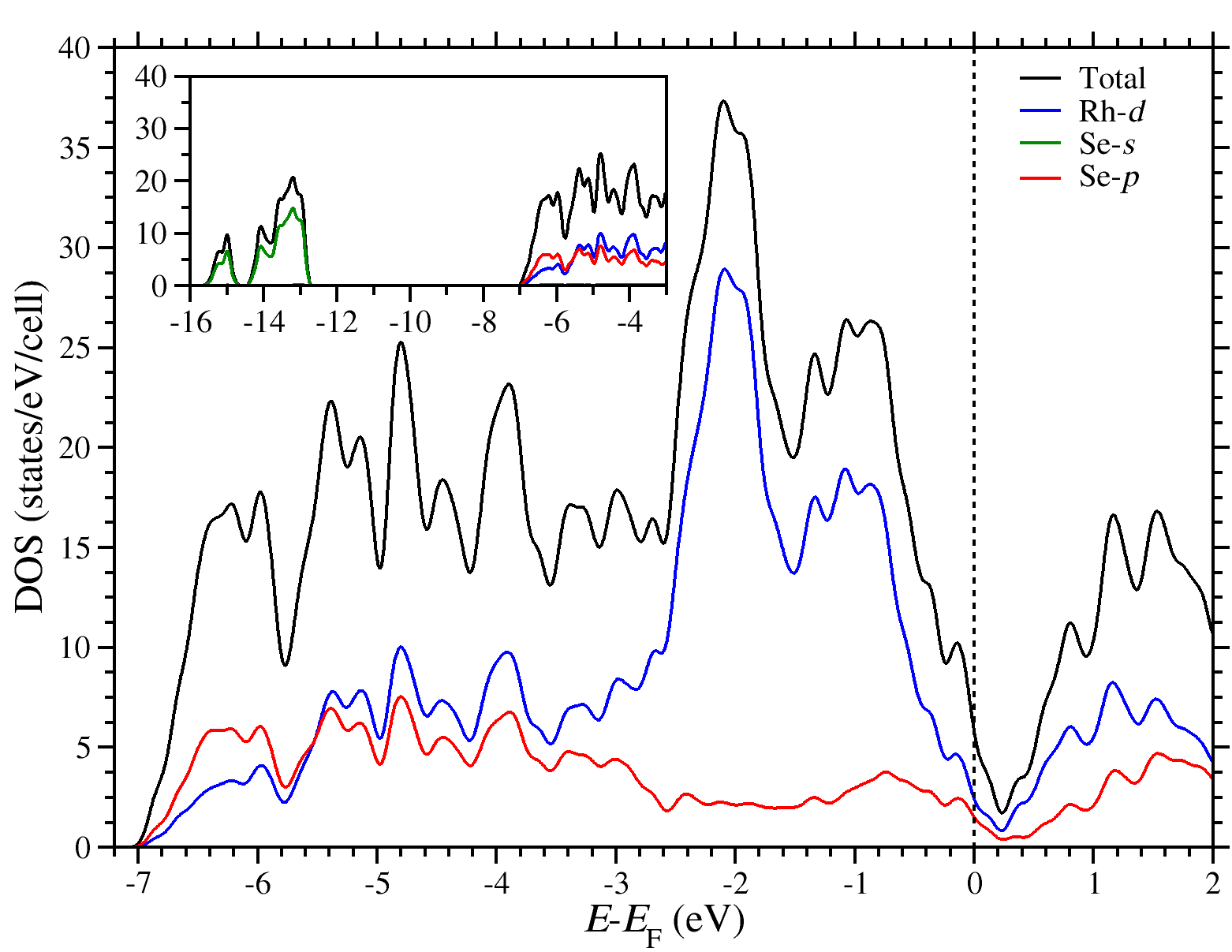}
    \caption{Rh$_3$Se$_4$ density of states. Black lines represent the total density of states, coloured lines represent the density of states projected on the $4d$ (blue) orbitals of the rhodium, $4s$ (green) and $4p$ (red) of the selenium.}
    \label{fig:Rh3Se4 PDOS}
\end{figure}

Total and projected density of states (DOS and PDOS, respectively) obtained for zaykovite are presented in Fig.~\ref{fig:Rh3Se4 PDOS}. As can be seen, Rh-$d$ and Se-$p$ states occupy and hybridize with each other over the entire interval above -7 eV.
The Se-$s$ states lie much deeper and a significant gap of about 5.5 eV is presented between the Se-$p$ and Se-$s$ state [Fig.~\ref{fig:Rh3Se4 PDOS}(inset)].
Interestingly, a pseudo-gap slightly above the Fermi level is observed. A similar pattern of density states with the pseudo-gap is seen in kingstonite as well \cite{Dieguez2009}.

Magnetic measurements of the sulfide counterpart Rh$_3$S$_4$ reveal the temperature independent paramagnetism with no increase in magnetic susceptibility at low temperatures~\cite{Beck2000}. Since selenium is in the same group with sulphur, zaykovite is expected to possess a similar magnetic state.

Our GGA calculations showed that zaykovite is non-magnetic. However, it is very well known that only account of Coulomb correlations can provide a correct description of magnetism in many transition metal compounds. Therefore, we carried out a series of GGA+$U$ calculations~\cite{Dudarev1998} with different (FM and AFM) initial magnetic structures at $U_\mathrm{eff}=U-J_\mathrm{H}$ values between 0 and 10 eV. The resulting ground state of the system was found non-magnetic up to $U_\mathrm{eff}=7$ eV. At unrealistically large $U_\mathrm{eff}>7$ eV, a ferromagnetic ground state is realized, similar to Rh$_3$S$_4$~\cite{Yu2015}. Therefore, one can expect that Rh$_3$Se$_4$ to be paramagnet. Interestingly, analysis of thermochemical data demonstrate that one can safely use $U=0$ for Rh$_3$S$_4$\cite{Yu2015}. In remain part of the paper we present results without taking into account of Hubbard correlation effects (we leave discussion of the importance of correlations for spectral properties for future studies, when corresponding experimental data will be available).

Finally, with the Stoner parameter calculated for Rh metal by Sigalas and Papaconstantopoulos \cite{Sigalas1994}, $I=0.309$ eV, and the density of states $D(E_\mathrm{F})= 0.641$ states/eV per atom, one can see that the Stoner criterion is not fulfilled:
\begin{eqnarray}
I\times D(E_\mathrm{F}) < 1
 \label{eq}
\end{eqnarray}
This explains, why the system prefers to remain paramagnetic.

\begin{figure}[t!]
\centering
\includegraphics[width=1\linewidth]{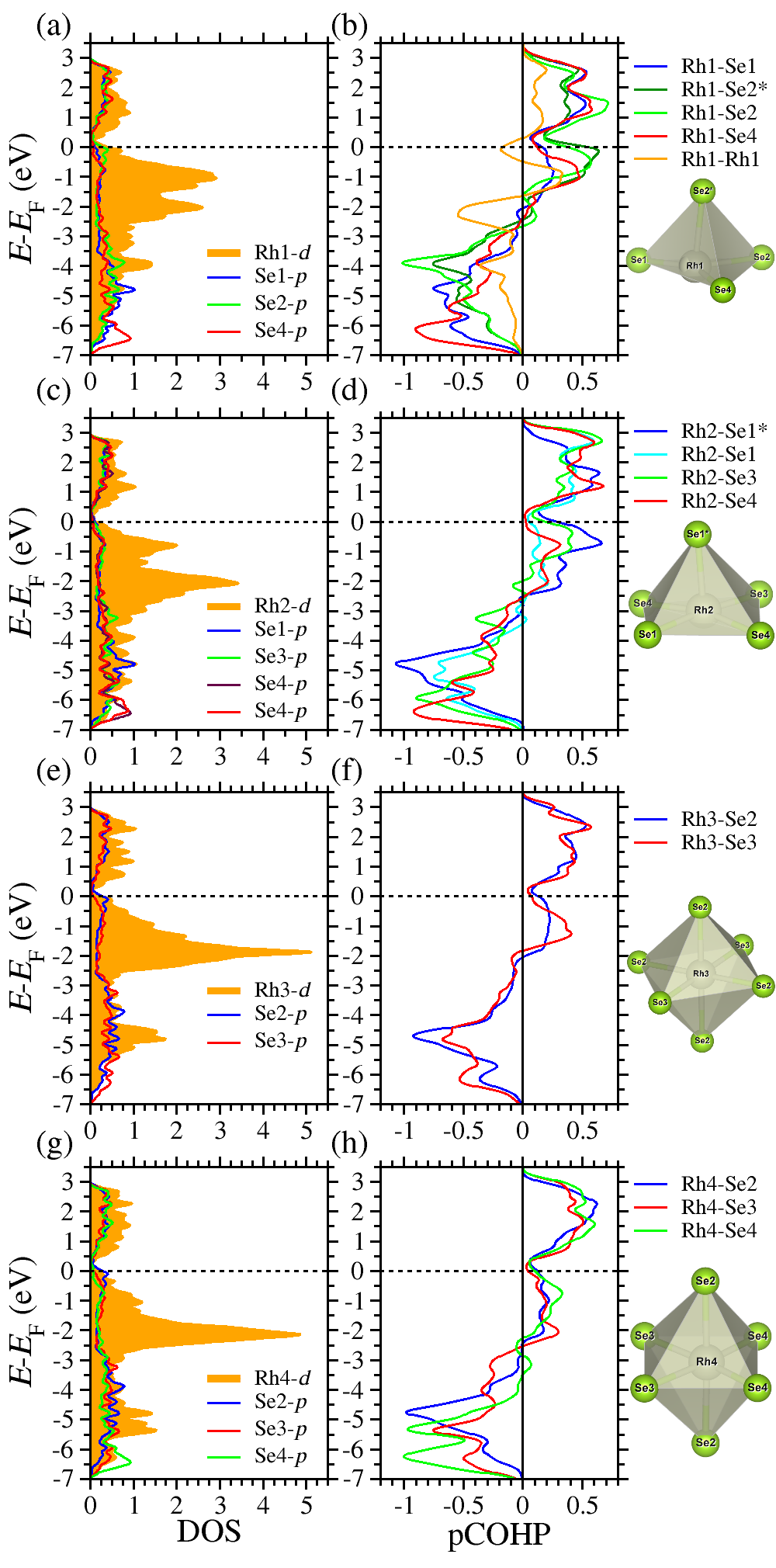}
    \caption{Density of states, DOS (in units of states/eV per polyhedron), and pCOHP curves} for Rh1 tetrahedra (a,b), Rh2 pyramids (c,d), Rh3 and Rh4 octahedra (e--h). For clarity, the structural elements are shown to the right of the plots.
    \label{fig:pCOHP}
\end{figure}

The projected COHP analysis \cite{Deringer2011,Maintz2013} is presented in Fig. \ref{fig:pCOHP}. For all variants of selenium environments, the Rh--Se bonding states (negative pCOHP) are residing from $-7$ to $\approx -2$ eV below the Fermi level, while antibonding combinations (positive pCOHP) are above $-2$ eV.

In order to get further insight into details of the chemical bonding, we calculated integrated pCOHP (IpCOHP) \cite{Dronskowski1993} and integrated crystal orbital bond index (ICOBI) \cite{Muller2021} (full list of -IpCOHPs and ICOBIs calculated for each interatomic bond are given in the supplementary material, Table~S1). First characterises the strength of the bond according to the principle: the more negative value, the stronger the bond is~\cite{Deringer2011,Maintz2016}. 
For the Rh--Se bonds, the bondstrength (-IpCOHP) vary from 1.87 to 2.6 eV depending on the bond length and the degree of distortion of the polyhedron, and for Rh1--Rh1 bond it equals to 0.82 eV, which is much weaker than for any Rh--Se bond. Integrated COBI is often used to characterize the degree of covalency of the bond under consideration: those with ICOBI close to 0 are typically ionic bonds, while ICOBI $\sim$ 1 is more specific for the covalent bonding. Average ICOBI for the Rh--Se bonds equals to 0.4, indicating a slight predominance of the ionic bond contribution over the covalent one. The ICOBI value for the Rh1--Rh1 bond is 0.26, indicating a strong ionic contribution.

\begin{figure}[t!]
\centering
\includegraphics[width=\linewidth]{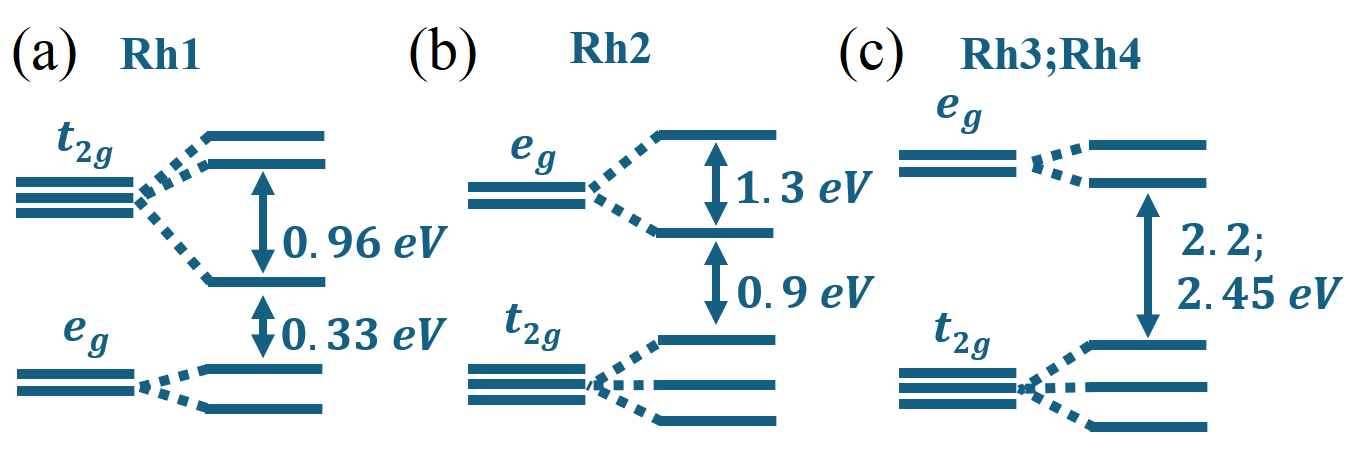}
    \caption{Crystal-field splitting as obtained by the Wannier function projection on Rh $4d$ states: (a) for Rh1 tetrahedra, (b) for Rh2 pyramids and (c) for Rh3, Rh4 octahedra.}
    \label{fig:Rh3Se4_Energy_levels}
\end{figure}

Thus, the study of pCOHP demonstrates that these are antibonding hybridized states of Rh-$4d$ and Se-$2p$ in the vicinity of the Fermi level. From the partial DOS plot presented in Fig.~\ref{fig:pCOHP}(a) one can also clearly see that $4d$ states of tetrahedral Rh1 are right below $E_F$. Naively, one might expect this is related to 
the fact that the $t_{2g}-e_g$ splitting in tetrahedra is 4/9 of what we have in octahedra (pyramids are cut octahedra), see e.g. Ref.~[\onlinecite{KHOMSKII-enc}]. Therefore, the pseudogap in DOS could be just a gap between low-lying ($t_{2g}$ in octahedra and pyramids, and $e_g$ in tetrahedra) and higher-lying ($e_{g}$ in octahedra and pyramids, and $t_{2g}$ in tetrahedra) states split by the crystal-field. However, the electron counting shows that this is not the case and there are 6 additional electrons, which can be distributed on the higher-lying $d$ levels ($e_{g}$ in octahedra and pyramids, and $t_{2g}$ in tetrahedra).

In order to have a realistic picture of $d$-level splitting, we used Maximally Localized Wannier Function (MLWF) technique \cite{Pizzi2020,Marzari1997,Souza2001}. The results presented in Fig.~\ref{fig:Rh3Se4_Energy_levels} show additional splitting of the high-energy states in tetrahedra and pyramids. Thus, for example, strong distortions of Rh1 tetrahedra (Rh1 is shifted nearly to one of the Se$_3$ faces) result in a strong splitting of the $t_{2g}$ states by 0.96 eV. In Rh2 pyramids $e_g$ splitting  equals 1.3 eV, shifting the $3z^2-r^2$ orbital downwards. Therefore, remaining 6 electrons are expected to occupy the split-off $xy$-orbital of tetrahedral Rh1 and the $3z^2-r^2$ orbital of pyramidal Rh2, so that the pseudogap is formed between these states and the higher-lying $xz/yz$ Rh1, $x^2-y^2$ Rh2 and $e_g$ orbitals of Rh3/Rh4. This agrees with the partial DOS states plotted in Fig.~\ref{fig:pCOHP}.

\begin{figure}[t!]
\centering
\includegraphics[width=\columnwidth]{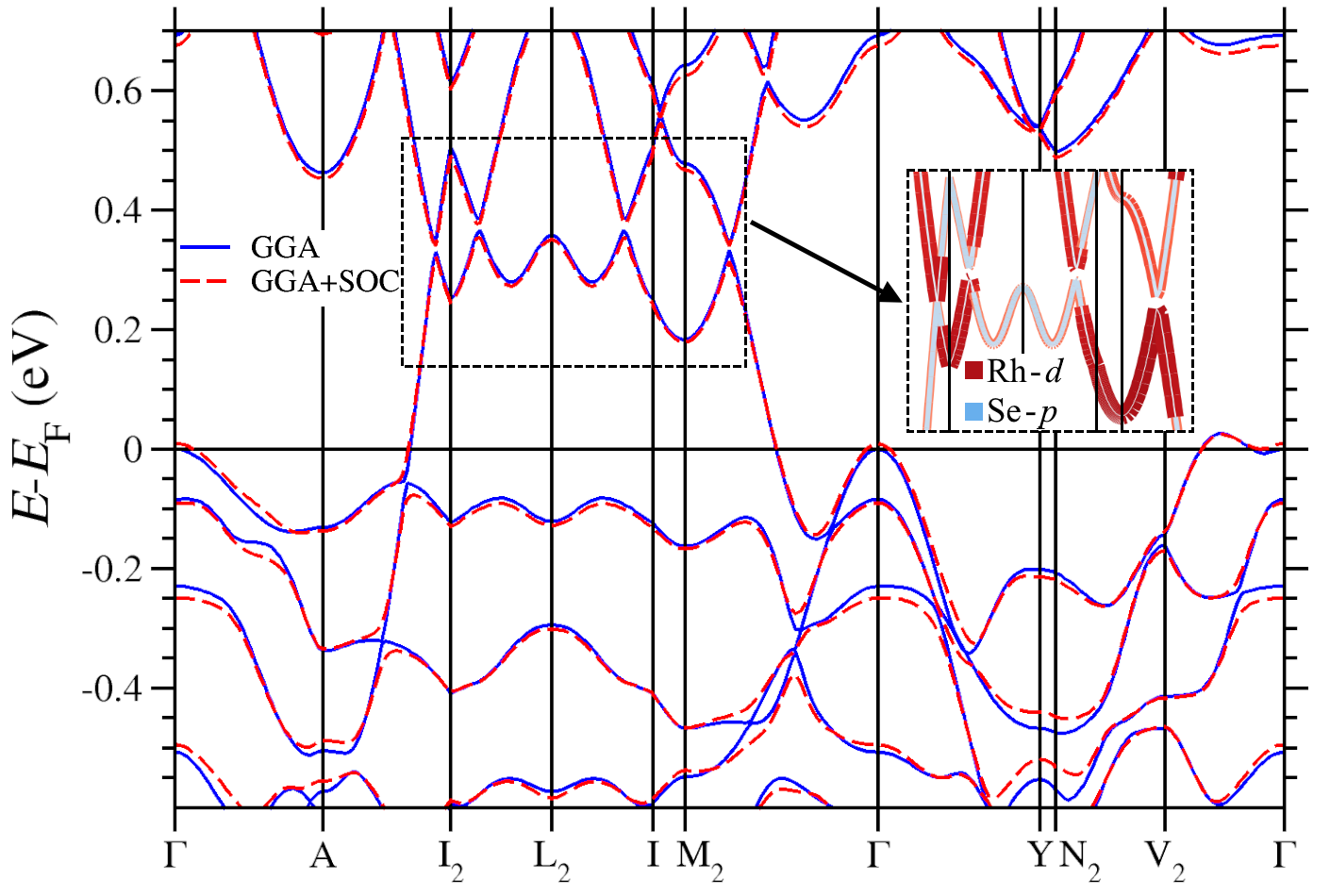}
    \caption{Band structure for Rh$_3$Se$_4$ in the primitive cell along the $k$-path suggested by the SeeK-path \cite{Hinuma2017,togo2024,Larsen2017}. Inset plotted using PyProcar \cite{Herath2020} shows the contributions of Rh-$d$ and Se-$p$ orbitals to bands in the vicinity of the gaps in the unoccupied part of the spectrum.}
    \label{fig:Rh3Se4 bands}
\end{figure}

Finally, we discuss details of the electronic dispersion in vicinity of the Fermi level shown in Fig. \ref{fig:Rh3Se4 bands}.  Gaps near the high-symmetry I$_2$, I, and M$_2$ points (\textit{A}-I$_2$-L$_2$ and L$_2$-I-M$_2$ paths) are observed, at energies of $\sim 0.350-0.375$ eV above the Fermi level. This corresponds to a small density of states on this interval in Fig. \ref{fig:Rh3Se4 PDOS}. The widths of the gaps vary from 7 to 19 meV for GGA and from 18 meV to 25 meV for GGA+SOC spectrum, and the Rh-$d$ contribution abruptly swaps across the gaps (Fig.~\ref{fig:Rh3Se4 bands}, inset), which could imply a possible band inversion. Although it should be noted that SOC does not significantly affect the dispersion of bands in the vicinity of the gaps. Despite the spectrum is metallic at the Fermi level we can consider its topological property assuming the bands below the gaps as the valence ones and given the presence of the inversion symmetry in the structure we can calculate the Z$_2$ topological invariant $(\nu_0;\nu_1\nu_2\nu_3)$ based on the products of the valence band Bloch wave functions parities in the TR-invariant momenta (TRIM) using the Fu-Kane formula~\cite{Fu2007PRB}. With the calculated parity products at TRIM (listed in Suppl.~Table~S2, see also Fig.~S1~(a)), the topological invariant is $(0;000)$, and hence the zaykovite is a topologically trivial system.

\begin{figure*}[t!]
\includegraphics[width=\textwidth]{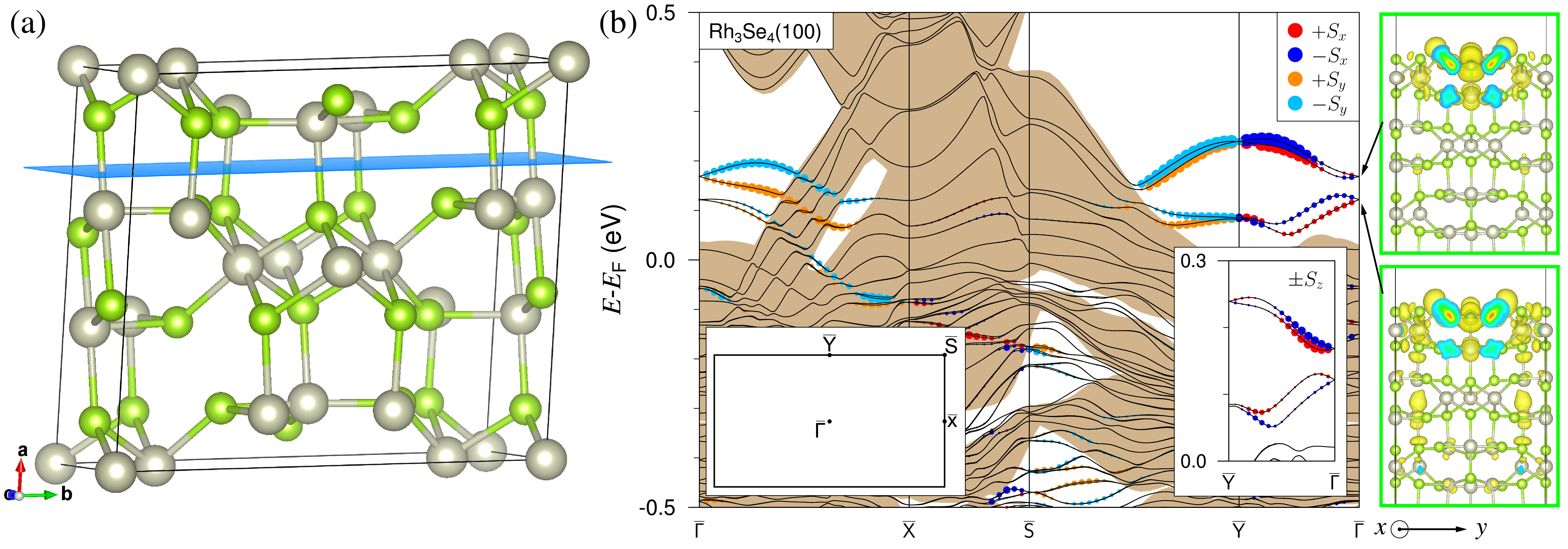}
    \caption{(a) Position of the $(100)$ cleavage plane (blue) within the conventional bulk cell. (b) Surface electronic structure calculated along high-symmetry directions of the surface BZ (left inset). Shaded area denotes bulk continuum states projected onto the $(100)$ surface. Red(orange)/blue(light blue) circles stand for $+S_\|$/$-S_\|$ spin projections of the surface states. Right inset shows the out-of-plane ($S_z$) spin component along $\bar{\mathrm Y}-\bar\Gamma$ direction, and right outsets demonstrate spatial localization of the surface states at the $\bar\Gamma$ point. }
    \label{fig:surf}
\end{figure*}

\subsection{Surface electronic structure}

To complement our analysis of the electronic properties of zaykovite, we will consider its surface electronic structure.
Despite the spin-orbit coupling does not  significantly affect the bulk electronic structure it can manifest itself in the surface spectrum via Rashba-type spin splitting \cite{Bihlmayer2022}.

Inspecting the crystal structure in detail, one can notice that the $(100)$ surface possesses the lowest density of Rh--Se bonds. Cleavage along this surface requires the breaking of six bonds (four from Rh1 tetrahedra and two from Rh2 pyramids) per $bc$ plane of the conventional cell [Fig.~\ref{fig:surf}(a)]. With known Rh--Se bond energies discussed above, the energy of such a cleavage can be estimated as 2.4 J/m$^2$ which is comparable with that in a well-known silicon \cite{Gleizer_PRL2014}. On the other hand, although the Rh$_3$Se$_4$ possesses the pseudo-layered structure with weakest bonds along the $[100]$ direction the obtained cleavage energy is about six times larger compared with that in van der Waals systems, for example, in graphite ($0.39\pm 0.02$ J/m$^2$, Ref.~[\onlinecite{Wang2015}]).

To calculate the surface electronic structure we constructed a slab of nine pseudo-layer thickness with a vacuum spacing of 15 \AA. The $x$ and $y$ directions in the slab coincide, respectively, with $c$ and $b$ vectors of the bulk cell, while the $z$ direction is perpendicular to the $bc$ plane. Surface relaxation leads to displacements of primarily the surface atoms, where the largest inward displacements are observed for topmost Rh atoms and outward displacements for topmost Se atoms, which, however, do not exceed 0.42 and 0.23 \AA, respectively.  

Figure~\ref{fig:surf}(b) demonstrates the Rh$_3$Se$_4$(100) surface band structure. In the metallic bulk band spectrum, there are wide gaps in the continuum states at the $\bar\Gamma$ and $\bar{\mathrm Y}$ points of the surface Brillouin zone (left inset) just above the Fermi level where two spin-split surface states reside. The upper one is entirely localized in the topmost pseudo-layer with minimum localization on the Rh$_6$ elements [see Fig.~\ref{fig:surf}(b), top-right outset]. Such localization makes this state quasi-one-dimensional, propagating in the form of stripes along the $x$-direction, which are largely isolated from each other in the $y$-direction. The latter leads to significant $\bar\Gamma-\bar {\mathrm X}$/$\bar\Gamma-\bar {\mathrm Y}$ anisotropy in the spin splitting. Along $\bar\Gamma-\bar {\mathrm X}$ the state demonstrates typical Rashba-type spin splitting with spins aligned completely in-plane and perpendicular to the $k_\|$ vector. Along $\bar\Gamma-\bar {\mathrm Y}$ the splitting between spin subbands is much smaller and in the vicinity the $\bar\Gamma$ point the spins are aligned along $z$ direction (right inset). The second surface state, lying closer to the bulk states demonstrates deeper penetration into the crystal, up to the third pseudo-layer [Fig.~\ref{fig:surf}(b), bottom-right outset], and a smaller anisotropy in the localization. This is reflected in its spin texture, which is predominantly in-plane both near the $\bar\Gamma$ and $\bar {\mathrm Y}$ points. Thus, despite the spin-orbit coupling has almost no effect on the bulk electronic spectrum (Fig.~\ref{fig:Rh3Se4 bands}) the emerging unoccupied surface states experience noticeable SOC-induced spin splitting.

\begin{table}[t!]
    \centering
    \caption{Unit cell parameters and $X$-Se bond lengths ($d_\mathrm{NN}$), where $X$=Ir, Pd or Pt, for the predicted selenides in relaxed unit cells using GGA-D3 method. Bonds with asterisks differ in length from their counterparts without an asterisk.}
    \renewcommand{\arraystretch}{1}
    \begin{ruledtabular}
    \begin{tabular}{cccc}
        Structure & Ir$_3$Se$_4$ & Pd$_3$Se$_4$ & Pt$_3$Se$_4$ \\ \midrule
        $a$ (\AA) & 10.940 & 11.293 & 11.303 \\ 
        $b$ (\AA) & 11.431 & 11.623 & 11.529 \\ 
        $c$ (\AA) & 6.562 & 6.775 & 6.843 \\
       \: $\alpha$ (deg) & 90 & 90  & 90  \\ 
       \: $\beta$ (deg) & 108.115 & 110.919 & 109.870 \\ 
       \: $\gamma$ (deg) & 90 & 90  & 90  \\ 
       \: $V$ (\AA$^3$) & 779.961 & 830.702 & 838.733 \\ \midrule
        \multicolumn{4}{c}{$d_\mathrm{NN}$ (\AA), tetrahedron environment} \\ \midrule
        $X$1-Se1 & 2.410 & 2.427 & 2.428 \\ 
        \, $X$1-Se2* & 2.516 & 2.573 & 2.552 \\ 
        $X$1-Se2 & 2.465 & 2.565 & 2.601 \\ 
        $X$1-Se4 & 2.369 & 2.443 & 2.446 \\ \midrule
        \multicolumn{4}{c}{$d_\mathrm{NN}$ (\AA), pyramidal environment} \\ \midrule
        \, $X$2-Se1* & 2.464 & 2.540 & 2.558 \\ 
        $X$2-Se1 & 2.492 & 2.583 & 2.575 \\ 
        $X$2-Se3 & 2.455 & 2.481 & 2.494 \\ 
        \qquad $X$2-Se4\,($\times 2$) & 2.393 & 2.456 & 2.430 \\ \midrule
        \multicolumn{4}{c}{$d_\mathrm{NN}$ (\AA), octahedral environment (type 1)} \\ \midrule
        \qquad $X$3-Se2\,($\times 4$) & 2.532 & 2.584 & 2.607 \\ 
        \qquad $X$3-Se3\,($\times 2$) & 2.485 & 2.535 & 2.547 \\ \midrule
        \multicolumn{4}{c}{$d_\mathrm{NN}$ (\AA), octahedral environment (type 2)} \\ \midrule
        \qquad $X$4-Se2\,($\times 2$) & 2.505 & 2.545 & 2.581 \\ 
        \qquad $X$4-Se3\,($\times 2$) & 2.472 & 2.534 & 2.554 \\ 
        \qquad $X$4-Se4\,($\times 2$) & 2.416 & 2.498 & 2.487 \\ 
    \end{tabular}
    \end{ruledtabular}
    \label{Tab: X3Se4struct}
\end{table}

\begin{table*}[t!]
    \centering
    \caption{Atomic coordinates for Ir$_3$Se$_4$, Pd$_3$Se$_4$, and Pt$_3$Se$_4$ as obtained by the relaxation of the crystal structure in GGA-D3.}
    \begin{ruledtabular}
    \begin{tabular}{lccclccclccc}
        \multicolumn{4}{c}{Ir$_3$Se$_4$} & \multicolumn{4}{c}{Pd$_3$Se$_4$} & \multicolumn{4}{c}{Pt$_3$Se$_4$}  \\ 
        \cmidrule(lr){1-4}
        \cmidrule(lr){5-8}
        \cmidrule(lr){9-12}
        Site & $x$ & $y$ & $z$ & Site & $x$ & $y$ & $z$ & Site & $x$ & $y$ & $z$  \\
        \cmidrule(lr){1-4}
        \cmidrule(lr){5-8}
        \cmidrule(lr){9-12}
        \,Ir1 (8\textit{j}) & 0.36795 & 0.14555 & 0.95574 &Pd1 (8\textit{j}) & 0.36750 & 0.14209 & 0.95418 &Pt1 (8\textit{j}) & 0.37079 & 0.13901 & 0.96108  \\ 
        \,Ir2 (4\textit{i}) & 0.35088 & 0 & 0.56343 &Pd2 (4\textit{i}) & 0.35119 & 0 & 0.56805 &Pt2 (4\textit{i}) & 0.34652 & 0 & 0.56403  \\ 
        \,Ir3 (2\textit{a}) & 0 & 0 & 0 &Pd3 (2\textit{a}) & 0 & 0 & 0 &Pt3 (2\textit{a}) & 0 & 0 & 0  \\ 
        \,Ir4 (4\textit{h}) & 0 & 0.16105 & 0.5 &Pd4 (4\textit{h}) & 0 & 0.16152 & 0.5 &Pt4 (4\textit{h}) & 0 & 0.16629 & 0.5  \\ 
       Se1 (4\textit{i}) & 0.41481 & 0 & 0.23503 &\;Se1 (4\textit{i}) & 0.41076 & 0 & 0.23367 &\:Se1 (4\textit{i}) & 0.41415 & 0 & 0.24321  \\ 
       Se2 (8\textit{j}) & 0.12908 & 0.15765 & 0.88913 &\;Se2 (8\textit{j}) & 0.12970 & 0.15540 & 0.89380 &\:Se2 (8\textit{j}) & 0.13057 & 0.15795 & 0.89102  \\ 
       Se3 (4\textit{i}) & 0.11696 & 0 & 0.39089 &\;Se3 (4\textit{i}) & 0.11691 & 0 & 0.39644 &\:Se3 (4\textit{i}) & 0.11347 & 0 & 0.39143  \\ 
       Se4 (8\textit{j}) & 0.35761 & 0.20799 & 0.60661 &\;Se4 (8\textit{j}) & 0.35565 & 0.21023 & 0.60652 &\:Se4 (8\textit{j}) & 0.35746 & 0.20868 & 0.61671  \\
    \end{tabular}
    \end{ruledtabular}
    \label{table:AtomCoords}
\end{table*}

\section{Prediction of related selenides}

As it has been mentioned above, the natural crystals of zaykovite contain inclusions of other transition metal elements (Ir, Pd or Pt), which are close to Rh in the periodic system. 
This suggests that there may be other selenides with the same crystal structure in which the Rh atoms are completely replaced by one of the atoms present as an impurity in the natural crystal.
To check this possibility, we calculated the dynamic and thermodynamic stabilities of these intended materials. 

The crystal structure optimization shows that Ir$_3$Se$_4$, Pd$_3$Se$_4$, and Pt$_3$Se$_4$ retain the same crystal structure as parent Rh$_3$Se$_4$. Corresponding lattice parameters and characteristic bond lengths are summarized in Table~\ref{Tab: X3Se4struct}, whereas atomic positions are given in Table \ref{table:AtomCoords}.

It is interesting to note that the equilibrium volume of Ir$_3$Se$_4$ is comparable to that of Rh$_3$Se$_4$, whereas the equilibrium volumes of Pd$_3$Se$_4$ and Pt$_3$Se$_4$ are significantly larger, although the ionic radii of Rh, Ir, Pd, and Pt are nearly the same~\cite{Shannon-76}. The origin of this behavior stems from the specific electronic structure of the selenides and a particular filling of the $d$ band which results in formation of the pseudogap in Ir$_3$Se$_4$ and Rh$_3$Se$_4$ close to the Fermi level.

More detailed analysis of the selenide polyhedra reveals additional distortions in Pd- and Pt-based compounds. While there is a noticeable bond-length difference of 0.05 \AA~ between $X1$-Se2 and $X1$-Se2* bonds for Ir$_3$Se$_4$ and Pt$_3$Se$_4$ (tetrahedra), see Table \ref{Tab: X3Se4struct}, in case of $X=\mathrm{Pd}$ they are almost the same. Moreover,  there is a compression of the pyramid along the plane (reduction of two equal lengths of Pt2-Se4 bonds, located opposite to each other). For Pd4 octahedra, the difference between pairs of bond lengths in the Pd4-Se3 and Pd4-Se4 planes is about two times smaller than for other selenides.

\begin{table}[b!]
    \centering
    \caption{Parameters for the equation of state for Ir$_3$Se$_4$, Pd$_3$Se$_4$, and Pt$_3$Se$_4$. $V_0$ stands for the equilibrium volume, $B_0$ is a bulk modulus.}
    \begin{ruledtabular}
    \begin{tabular}{lccc}
       & Ir$_3$Se$_4$ & Pd$_3$Se$_4$ & Pt$_3$Se$_4$ \\ \midrule
        $V_0$ (\AA${^3}$) & 401.85 & 431.73 & 434.75 \\ 
        $B_0$ (GPa)   & 140.46 & 90.33 & 108.03 \\ 
    \end{tabular}
    \end{ruledtabular}
    \label{table:EOS}
\end{table}
\begin{figure*}[t!]
\centering
\includegraphics[width=\linewidth]{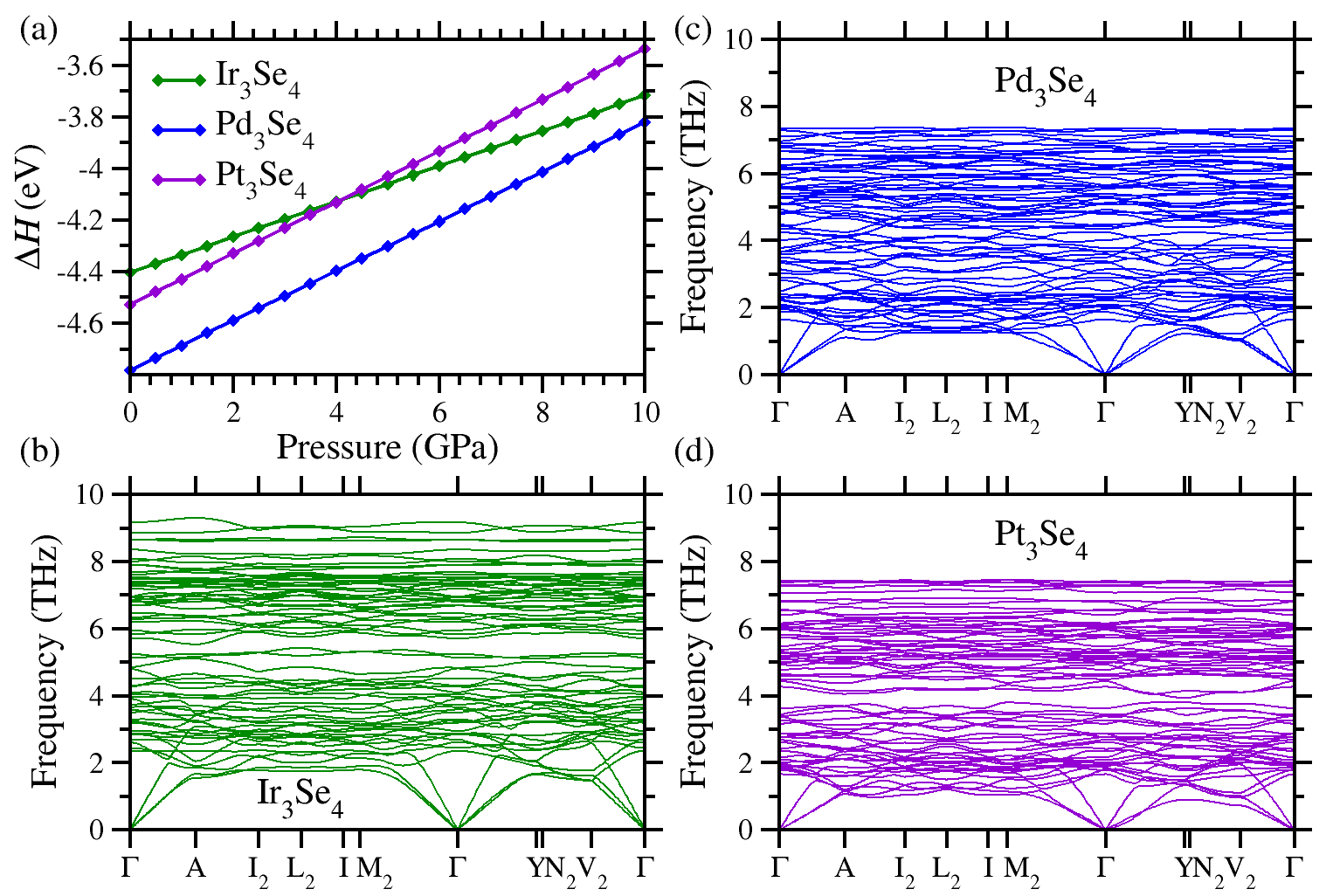}
    \caption{(a) Relative enthalpies of the predicted selenides. Phonon band spectra of (b) Ir$_3$Se$_4$, (c) Pd$_3$Se$_4$, and (d) Pt$_3$Se$_4$. Three bands showing zero frequency at $\Gamma$ point are the acoustic modes, and the rest are optical modes. }
    \label{fig:Stability}
\end{figure*}
\begin{figure*}[t!]
\centering
\includegraphics[width=\linewidth]{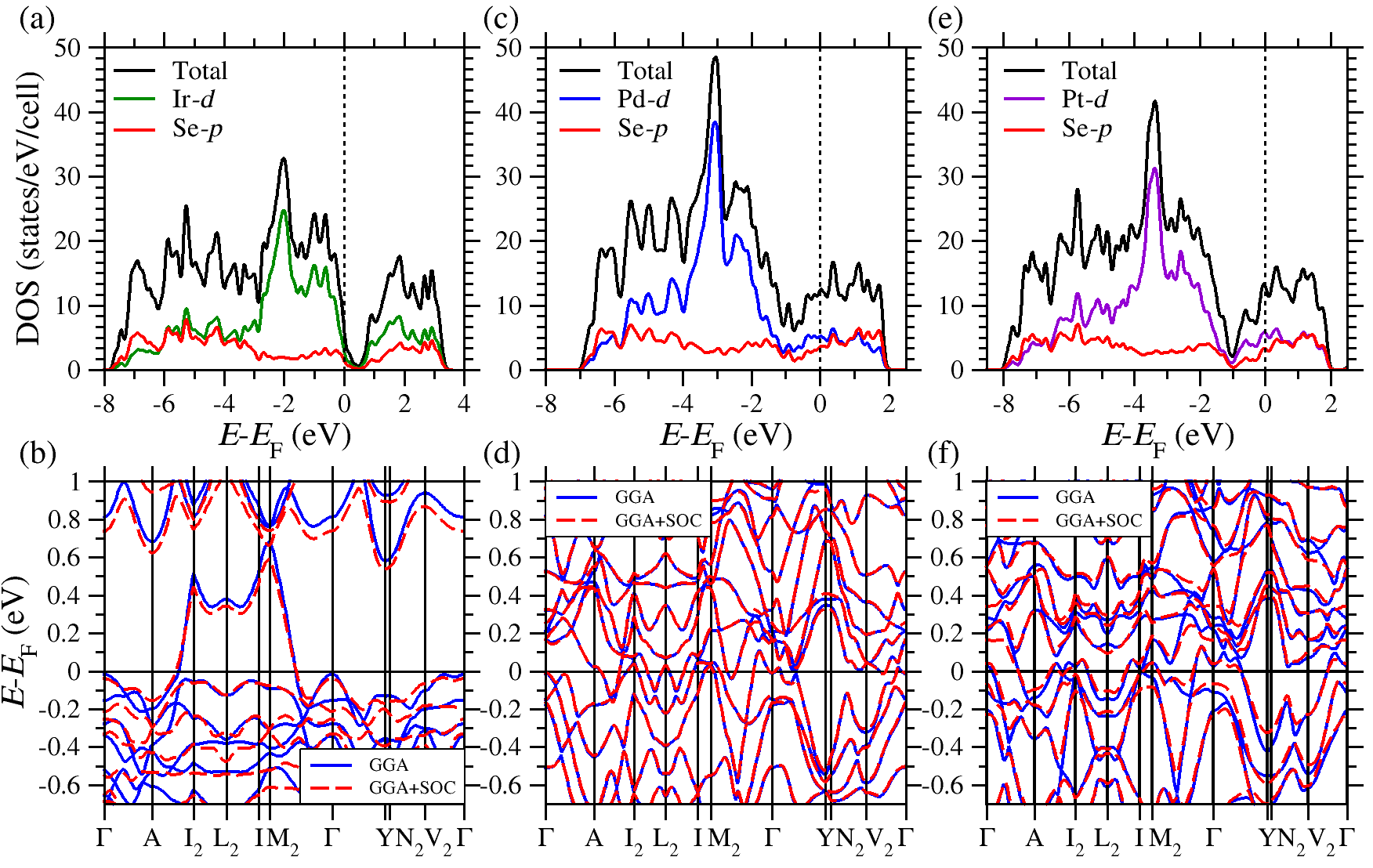}
    \caption{Density of states and band structure of Ir$_3$Se$_4$ (a,b); Pd$_3$Se$_4$ (c,d) and Pt$_3$Se$_4$ (e,f). Fermi level is set to zero. 
    }
    \label{fig:DOS_bands}
\end{figure*}

To check the thermodynamic stability, first the total energy dependence on the primitive cell volume for selenides and pure elements (Ir, Pd, Pt and Se) was obtained from series of calculations, where the volume interval varied between about -10~\% and +10~\% from its equilibrium value. The energy--volume curves for Ir$_3$Se$_4$, Pd$_3$Se$_4$, and Pt$_3$Se$_4$ compounds are given in Suppl. Figs.~S2~(a--c).
Further, the parameters of the equation of state (EOS) such as the equilibrium volume $V_0$, and the bulk modulus $B_0$ were extracted by fitting the calculated energies versus volume to the third order Birch–Murnaghan EOS~\cite{Birch1947} and presented in Table \ref{table:EOS}.
The thermodynamic stability was investigated by comparing the enthalpies of $X_3$Se$_4$ ($X=\mathrm{Ir,Pd, Pt}$) and its constituent pure elements, which are shown in Fig. \ref{fig:Stability}(a). It can be seen that enthalpy of formation, $\Delta H$, is negative in a wide range of pressure, so the predicted compounds are thermodynamically stable.

To clarify the issue of the dynamic stability of the predicted selenides, we performed the first-principles phonon calculations. As shown in Fig.~\ref{fig:Stability}~(b--d), there are no imaginary modes in the phonon spectra of all suggested compounds, meaning they are dynamically stable. The maximum phonon frequencies for Ir$_3$Se$_4$ reach 9 THz (similar to Rh$_3$Se$_4$, see the phonon density of states (phDOS) in Fig.~S3~(a)), while those for Pd$_3$Se$_4$ and Pt$_3$Se$_4$ are about 7.5 THz. This is directly related to a smaller bulk modulus in case $X=\mathrm{Pd, Pt}$.

Another characteristic feature of Ir$_3$Se$_4$ and Pt$_3$Se$_4$ phonon spectra is a clear separation of low-frequency vibration modes of heavy Ir and Pt and high-frequency selenium modes with distinct gaps at $\approx 5.5$ and $\approx 4$ THz, respectively (Figs.~\ref{fig:Stability}~(b,d), see also phDOS in the Suppl. Figs.~S3~(b,d)].

The densities of electronic states and bulk band spectra of $X_3$Se$_4$ compounds are presented in Fig.~\ref{fig:DOS_bands}. Similar to Rh$_3$Se$_4$, the DOS of the compound with isoelectronic Ir possesses the pseudogap above the Fermi level [Fig.~\ref{fig:DOS_bands}~(a)]. This pseudogap in DOS at $\approx 0.7$~eV above $E_\mathrm{F}$ comes from indirect (negative) $\mathrm{M}_2-\mathrm{Y}$ gap of $-0.05$~eV in the band spectrum  [Fig.~\ref{fig:DOS_bands}~(b)] which, unlike Rh$_3$Se$_4$, does not demonstrate an inversion of $d$-states at the I$_2$ and M$_2$ points. The SOC affects the band structure of Ir$_3$Se$_4$ stronger than in the case of Rh$_3$Se$_4$, because the strength of the spin-orbit coupling, characterized by $\lambda$ for $5d$ transition metals is larger than for $4d$. Typically, $\lambda_{4d}\sim~0.1-0.2$ eV~\cite{Dunn1961}, and $\lambda_{5d}\sim~0.3-0.5$ eV~\cite{Yuan2017}. Despite the SOC causes a stronger change in the electronic spectrum of Ir$_3$Se$_4$, the parity calculations (Table~S1) show that the compound, like zaykovite, has a trivial topological phase. On the other hand, in the iridium selenide one can expect more significant spin splitting in the surface states.

Pd and Pt atoms are in the next group of the periodic table and contain one more electron on the $d$ orbital compared to Rh and Ir. This leads to a shift of $t_{2g}$ and $e_g$ states deeper by $\sim 1$ eV. As a result, the spectra of Pd$_3$Se$_4$ and Pt$_3$Se$_4$ are entirely metallic in the vicinity of the Fermi level [Fig.~\ref{fig:DOS_bands}~(d,f)] and the pseudogap in Pt$_3$Se$_4$ and Pd$_3$Se$_4$ DOSs also shifts  by $\sim 1$ eV below the Fermi level [Fig.~\ref{fig:DOS_bands}~(c,e)], although in the latter case it is less pronounced. Finally, note that the Stoner criterion (\ref{eq}) for $X_3$Se$_4$ compounds is not fulfilled either (see Table~\ref{table:StonerCriterion}) and all of them are nonmagnetic like Rh$_3$Se$_4$.

As noted above, natural samples of zaykovite can comprise significant amounts of $X$ metal impurities. In particular, it can contain up to 19~\% of platinum \cite{Belogub2023}. Having constructed the (Rh$_{0.81}$Pt$_{0.19}$)$_3$Se$_4$ system, the crystal structure of which was calculated from the equilibrium structures of Rh$_3$Se$_4$ and Pt$_3$Se$_4$ following the Vegard's law, we simulated the Rh-Pt site intermix using the virtual crystal approximation (VCA)~\cite{VCA}. This Pt admixture leads to the shift the pseudogap from $\approx +0.35$~eV in pristine zaykovite to $\approx -0.14$~eV in the Pt-doped case (see Suppl. Fig.~S4). It is obvious that with a smaller doping $x$ in the (Rh$_{1-x}$Pt(Pd)$_{x}$)$_3$Se$_4$ samples, this pseudogap can be strictly at the Fermi level. The presence of sulfur impurity in natural minerals can also partially modify the electronic structure.

\begin{table}[h!]
    \centering
    \caption{Stoner criterion parameters for Ir$_3$Se$_4$, Pd$_3$Se$_4$, and Pt$_3$Se$_4$.}
    \begin{ruledtabular}
    \begin{tabular}{lccc}
       & Ir$_3$Se$_4$ & Pd$_3$Se$_4$ & Pt$_3$Se$_4$ \\ \midrule
        $I$ (eV)~\cite{Sigalas1994} & 0.295 & 0.313 & 0.299 \\ 
        $D(E_\mathrm{F})$ (states/eV/atom)   & 0.721 & 1.362 & 1.453 \\ 
    \end{tabular}
    \end{ruledtabular}
    \label{table:StonerCriterion}
\end{table}

\section{Conclusions}

In summary, in this work we have scrutinized the bulk and surface electronic structure of recently discovered mineral zaykovite having chemical formula Rh$_3$Se$_4$. We have shown that the inclusion of dispersion force corrections is important for an accurate description of the equilibrium crystal structure. The compound was determined to be paramagnetic semimetal, both without and with spin-orbit coupling (SOC) included in the calculations, with pseudogap in the electronic spectrum just above the Fermi level. The inspection of topological properties of Rh$_3$Se$_4$ shows the absence of the non-trivial band topology. Analysis of the hybridization between orbitals of rhodium and selenium demonstrated that the bonding states lie deep in the occupied part of the spectrum, while antibonding Rh-Se states define electronic structure in the vicinity of $E_\mathrm{F}$. The Rh--Se bonds demonstrate slight predominance of the ionic  bonding over the covalent one. We have shown that the $(100)$ surface, which has the lowest density of Rh--Se bonds, has a relatively low cleavage energy, only about six times larger compared with graphite. The surface supports the localized states in the local gap above $E_\mathrm{F}$ demonstrating noticeable anisotropy in their spatial localization,  band dispersion, and spin-orbit coupling induced spin splitting.

Additionally, we also predicted the dynamic and thermodynamic stability of $X_3\mathrm{Se}_4$ $(X=\mathrm{Ir, Pd, Pt})$ materials with the same crystal structure. We presented their equilibrium crystal structure parameters and identified the features of electronic properties depending on the $X$ metal.

\section*{Supplementary material}
The supplementary material contains chemical bonding data such as -IpCOHP and ICOBI; wave function parity products $\delta_i$ at time-reversal invariant momenta; dependence of total energy on the volume of primitive cell for $X_3\mathrm{Se}_4$ ($X = \mathrm{Ir, Pd, Pt}$); phonon density of states analysis for the compounds presented in the main paper; crystal structure and band spectrum of $(\mathrm{Rh}_{1-x}\mathrm{Pt}_x)_3\mathrm{Se}_4$.

\begin{acknowledgments}
We are grateful to E.V. Belogub who paid our attention on zaykovite, to M.M. Otrokov for
discussions of topological properties of Rh$_3$Se$_4$ and related materials, and to A.E. Lebedeva for participating in the early stages of the study. Work of Yekaterinburg's group was supported by the Ministry of Science and Higher Education of the Russian Federation through the ``Quantum'' program (No 122021000038-7). Phonon computations were performed on the Uran supercomputer at the IMM UB RAS. S.V.E. acknowledges the support by the Government research assignment for ISPMS SB RAS, project FWRW-2022-0001. The calculations were partly performed using the equipment of Shared Resource Center "Far Eastern Computing Resource" IACP FEB RAS (https://cc.dvo.ru).
\end{acknowledgments}

\section*{Author declarations}
\subsection*{Conflict of interest}
The authors have no conflicts to disclose.
\subsection*{Author contributions}
\textbf{Leonid S. Taran}:  Calculations (equal); Investigation (equal); Visualization (equal); Writing -- Original Draft Preparation (lead). \textbf{Sergey V. Eremeev}:  Calculations (equal); Analysis (lead); Investigation (equal); Visualization (equal); Writing -- Review \& Editing (equal); Supervision (equal). \textbf{Sergey V. Streltsov}: Idea (lead), Project Administration (lead); Writing -- Review \& Editing (equal); Supervision (equal).

\section*{DATA AVAILABILITY}
The data that support the findings of this study are available within the article and its supplementary material.
\appendix

\bibliography{Main}

\end{document}


\makeatletter
\def\@email#1#2{%
 \endgroup
 \patchcmd{\titleblock@produce}
  {\frontmatter@RRAPformat}
  {\frontmatter@RRAPformat{\produce@RRAP{*#1\href{mailto:#2}{#2}}}\frontmatter@RRAPformat}
  {}{}
}%
\makeatother
\title{Supplementary material for: \\
Electronic structure of zaykovite Rh$_3$Se$_4$, prediction and analysis of physical properties of related materials: Pd$_3$Se$_4$, Ir$_3$Se$_4$, and Pt$_3$Se$_4$}

\author{Leonid S. Taran}
\email{leonidtaran97@gmail.com}
\affiliation{M. N. Mikheev Institute of Metal Physics, Ural Branch of Russian Academy of Sciences, 620137 Yekaterinburg, Russia}

\author{Sergey V. Eremeev}%
\affiliation{Institute of Strength Physics and Materials Science of Siberian Branch of Russian Academy of Sciences, 634055, Tomsk, Russia}

\author{Sergey V. Streltsov}
\affiliation{M. N. Mikheev Institute of Metal Physics, Ural Branch of Russian Academy of Sciences, 620137 Yekaterinburg, Russia}

\date{\today}

\maketitle

\section{Chemical bonding}

Table \ref{table:ChemBond} shows negative values of IpCOHP and ICOBI values for zaykovite. As can be seen, all ICOBI values range from 0.31 to 0.45 for Rh--Se bonds which indicating a slight predominance of the ionic bond
contribution over the covalent one, how it was discussed in the main paper.
\begin{table}[h!]
    \centering
    \caption{-IpCOHP and ICOBI for interatomic bonds in Rh$_3$Se$_4$. Bonds with asterisks differ in length from their counterparts without an asterisk.}
    \begin{ruledtabular}
    \begin{tabular}{lll}
      Bond & -IpCOHP & ICOBI \\ \midrule
        Rh1-Se1  & 2.59714 & 0.44451 \\ 
        Rh1-Se2* & 1.86910 & 0.33217 \\ 
        Rh1-Se2  & 2.20879 & 0.42124 \\
        Rh1-Se4  & 2.58971 & 0.44057 \\
        Rh1-Rh1  & 0.81832 & 0.26470 \\
        Rh2-Se1* & 1.87002 & 0.31148 \\
        Rh2-Se1  & 2.13366 & 0.40774 \\
        Rh2-Se3  & 2.26927 & 0.38480 \\
        Rh2-Se4  & 2.56450 & 0.43328 \\
        Rh3-Se2  & 2.05782 & 0.38796 \\
        Rh3-Se3  & 2.13914 & 0.39151 \\
        Rh4-Se2  & 2.09632 & 0.39562 \\
        Rh4-Se3  & 2.18005 & 0.40914 \\
        Rh4-Se4  & 2.45826 & 0.39567 \\
    \end{tabular}
    \end{ruledtabular}
    \label{table:ChemBond}
\end{table}

\section{Topological invariants}

We calculate four independent Z$_2$ topological indices $(\nu_0;\nu_1\nu_2\nu_3)$ based on the parity products in the TR-invariant momenta (TRIM) $\Gamma_{i}$ using the Fu-Kane formula~\cite{Fu2007PRB}.

The $\nu_0$ is expressed as the product over
all eight points,
\begin{equation}
    (-1)^{\nu_0}=\prod^8_{i=1}\delta_i.
\end{equation}

The other three indices are given by products of four $\delta_i$’s:
\begin{equation}
    (-1)^{\nu_k}=\prod_{n_{k=1};n_{j\neq k}=0,1}\delta_{i=(n_1n_2n_3)}.
\end{equation}

\begin{figure*}
    \centering
    \includegraphics[width = 1\linewidth]{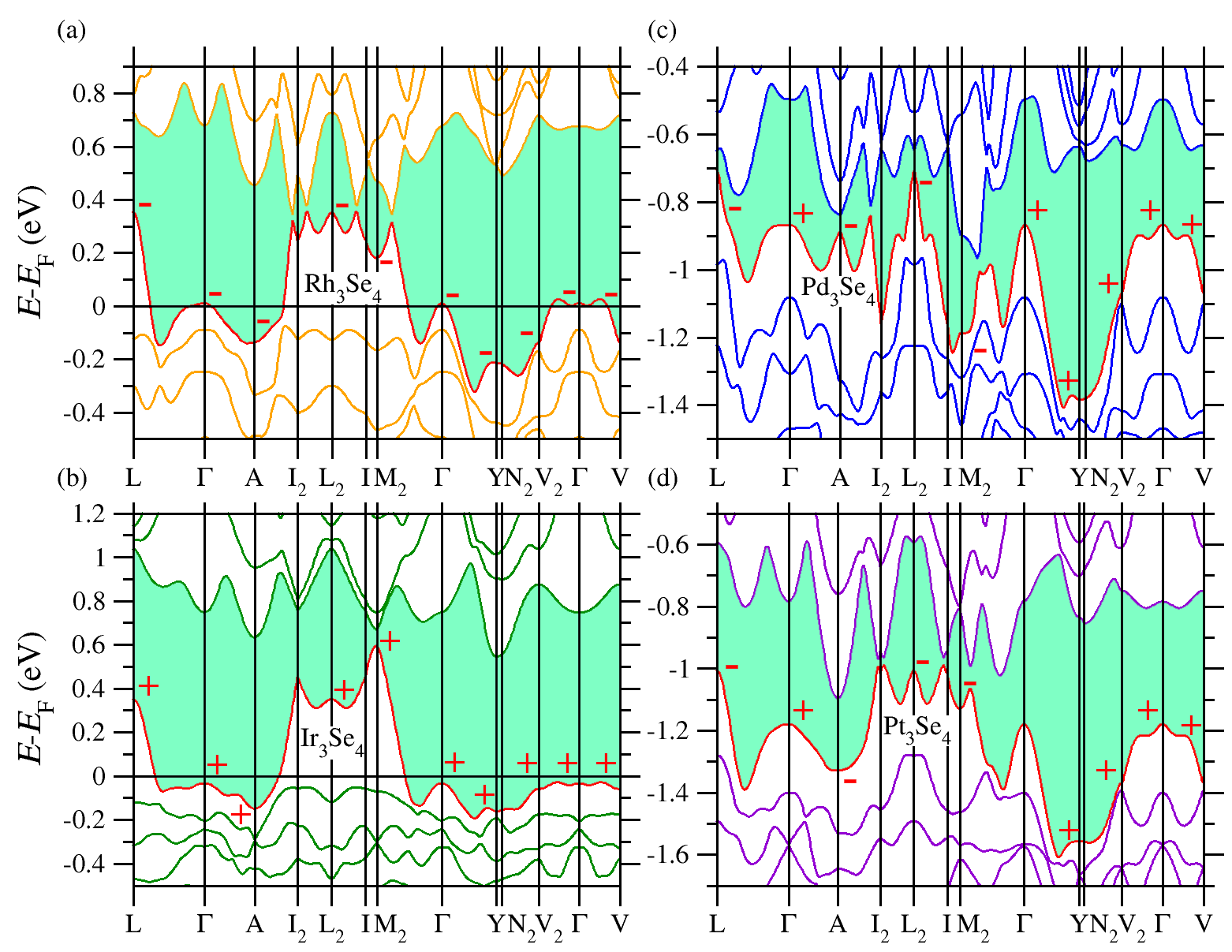}
    \caption{Band structure and parity products $\delta_{i}=\pm 1$ taken to band \#154 (marked in red color) for (a) Rh$_3$Se$_4$, (b) Ir$_3$Se$_4$, (c) Pd$_3$Se$_4$ and (d) Pt$_3$Se$_4$. Mint-green fill shows the area of the indirect gap.}
    \label{fig:TRIMbands}
\end{figure*}

The calculated $\delta_i$’s are given in Table~\ref{tab:trim} and shown in Fig. \ref{fig:TRIMbands}.

\begin{table}[h!]
    \centering
        \caption{Wave function parity products $\delta_{i}$ (taken to band \#154) in the time reversal invariant momenta (TRIM) points $\Gamma_{i=(n_1n_2n_3)}$.}
    \begin{tabular}{r | c c c c c c c c}
 $(n_1n_2n_3)$ & $(000)$ & $(111)$ & $(100)$ & $(010)$ & $(001)$ & $(110)$ & $(101)$ &  $(011)$ \\ 
  & $\Gamma$ & M$_2$ & V & V$_2$ & A & Y & L &  L$_2$ \\ \midrule
Rh$_3$Se$_4$ & $-$            & $-$          & $-$          & $-$          & $-$          & $-$          & $-$          & $-$ \\
Ir$_3$Se$_4$ & $+$            & $+$          & $+$          & $+$          & $+$          & $+$          & $+$          & $+$ \\
Pd$_3$Se$_4$ & $+$            & $-$          & $+$          & $+$          & $-$          & $+$          & $-$          & $-$ \\
Pt$_3$Se$_4$ & $+$            & $-$          & $+$          & $+$          & $-$          & $+$          & $-$          & $-$

    \end{tabular}
    \label{tab:trim}
\end{table}

\section{$E(V)$ curves}

\begin{figure*}
   \centering
    \includegraphics[width = 1\linewidth]{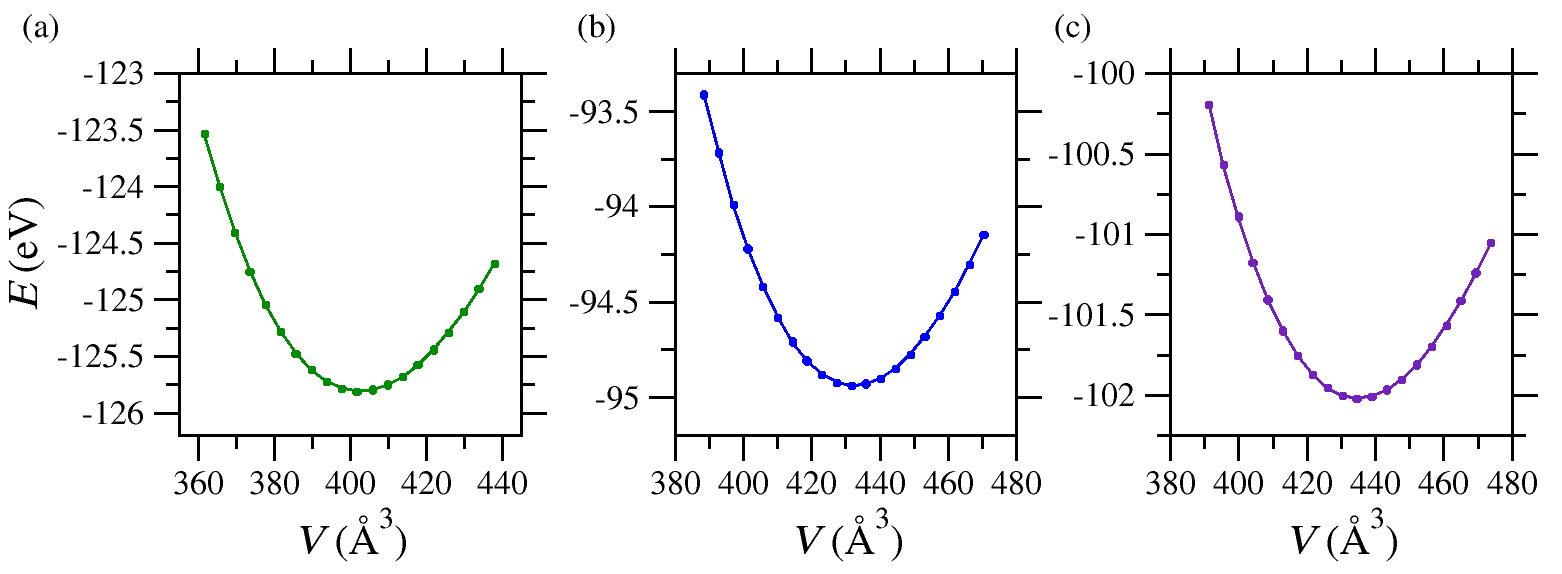}
    \caption{$E(V)$ curves for (a) Ir$_3$Se$_4$, (b) Pd$_3$Se$_4$ and (c) Pt$_3$Se$_4$.}
    \label{fig:Energy_curves}
\end{figure*}

Dependence of total energy on the volume of primitive cell for $X_3\mathrm{Se}_4$ ($X= \mathrm{Ir, Pd, Pt}$) are given in Fig.~\ref{fig:Energy_curves}.

\section{Phonon Density of States}

\begin{figure*}
    \centering
    \includegraphics[width = \linewidth]{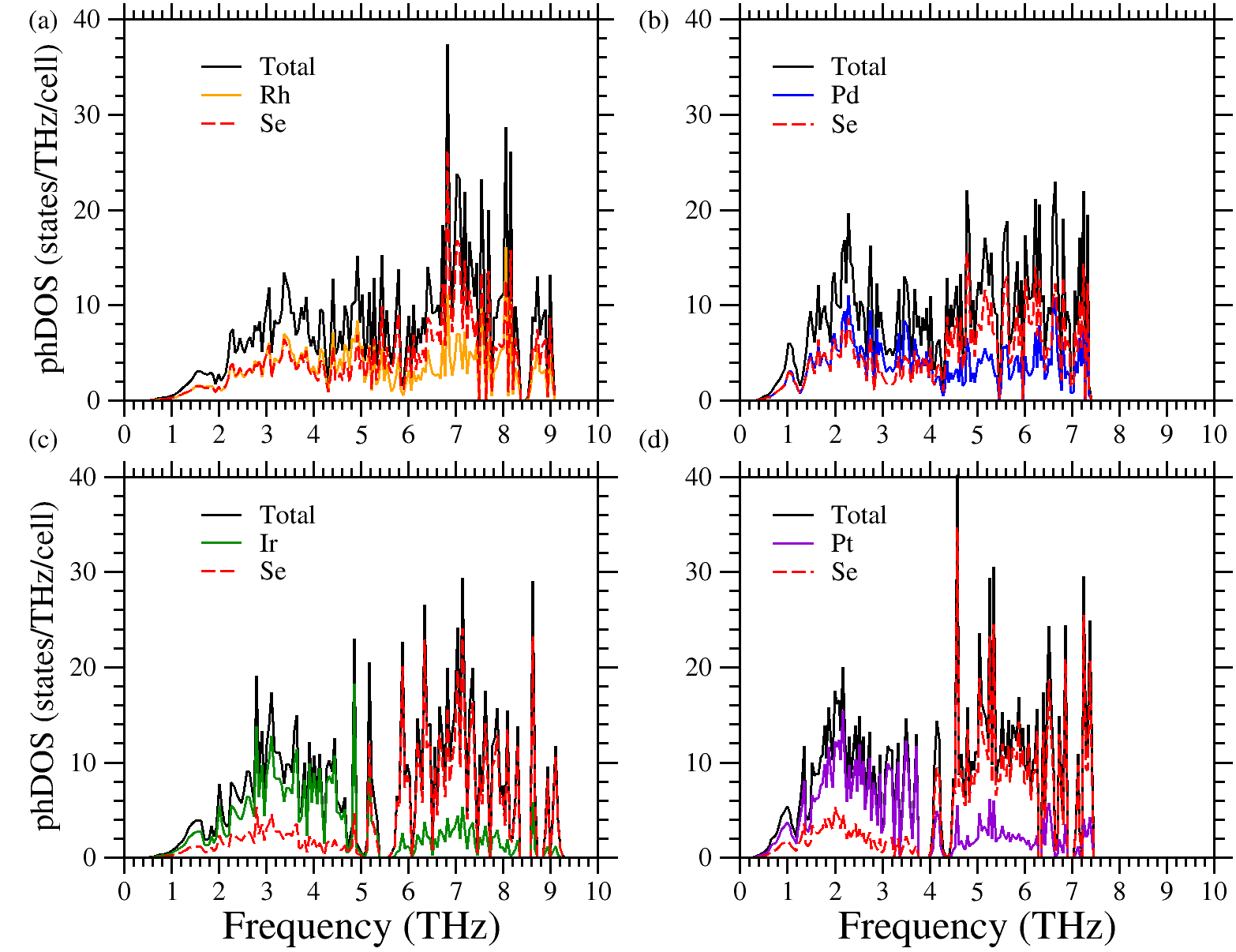}
    \caption{Phonon density of states phDOS of (a) Rh$_3$Se$_4$, (b) Ir$_3$Se$_4$, (c) Pd$_3$Se$_4$ and (d) Pt$_3$Se$_4$.}
    \label{fig:Phonon_PDOS}
\end{figure*}

Plotted in Fig. \ref{fig:Phonon_PDOS} total and partial phonon densities of states show that for compounds with $5d$ atoms Ir$_3$Se$_4$ and Pt$_3$Se$_4$  there is a clear separation of partial Se and metal atom phDOS: vibration states of the lightweight selenium atoms are predominantly located at the frequencies above, while those of heavy Rh and Ir states below 5.5 and 4 THz, respectively. 
In the case of Rh$_3$Se$_4$ and Pd$_3$Se$_4$ the picture is not so unambiguous: the partial density of states of selenium and $4d$ atoms equals in the low frequency region,  up to 3--4 THz. Unlike the $5d$-based compounds there are no gaps in the middle region and in the high frequency region the partial phDOS of $4d$ atoms dominate however, the Se contribution is also substantial: it is only 1.5--2 times smaller. 

\section{Electronic structure of $(\mathrm{Rh}_{1-x}\mathrm{Pt}_x)_3\mathrm{Se}_4$}

\begin{figure*}
    \centering
    \includegraphics[width = \linewidth]{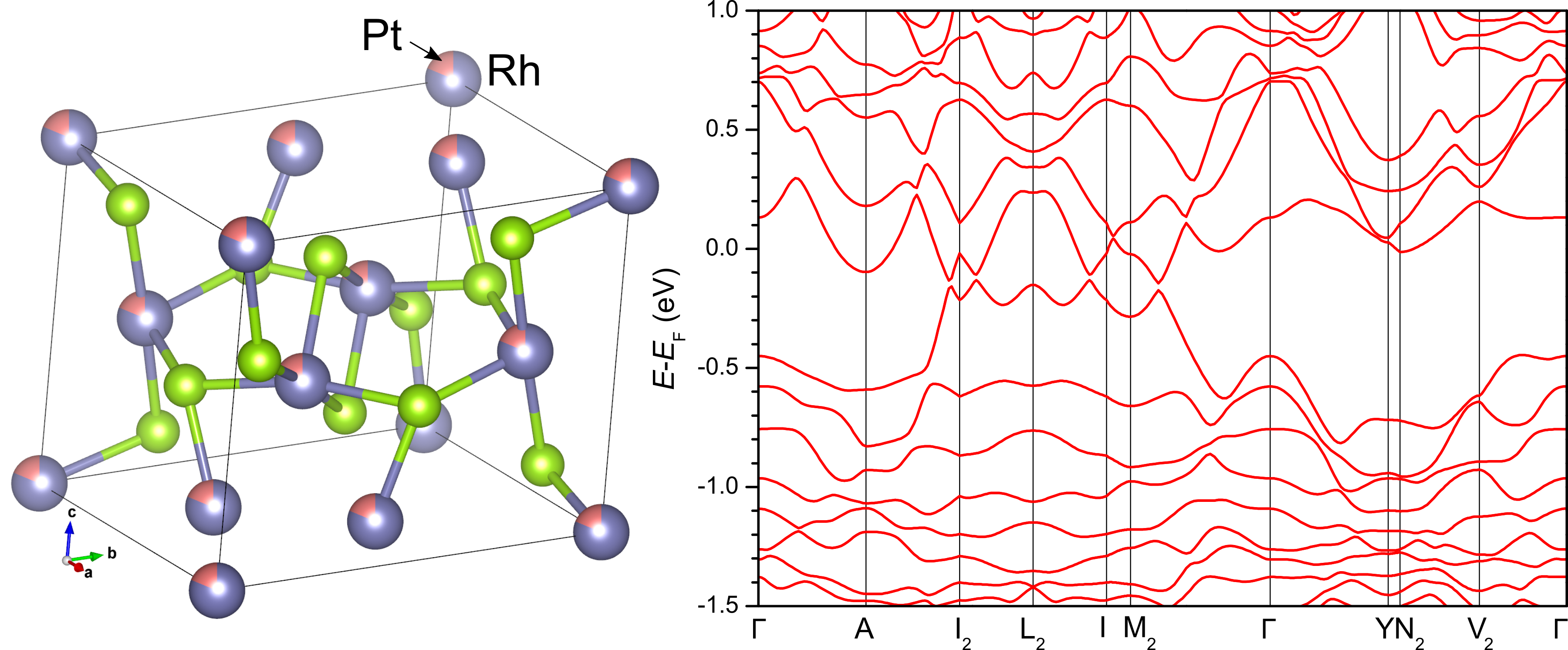}
    \caption{Crystal structure and band spectrum of (Rh$_{0.81}$Pt$_{0.19}$)$_3$Se$_4$.}
    \label{fig:Rh-Pt}
\end{figure*}

Primitive cell of $(\mathrm{Rh}_{1-x}\mathrm{Pt}_x)_3\mathrm{Se}_4$ compound with Pt alloying of concentration $x=0.19$ evenly distributed over the Rh sublattice and its electronic spectrum are presented in Fig.~\ref{fig:Rh-Pt}.

\bibliography{Main}